\documentclass[11pt]{article} 
\usepackage{mystyle-new}
\usepackage{authblk}
\usepackage[T1]{fontenc}
\usepackage{epsfig,amsmath} 
\usepackage{hepnames,hepunits}
\usepackage{hyperref}
\usepackage{color}
\usepackage{graphicx}
\definecolor{red}{rgb}{1,0,0}
\def\lesssim{\ \hbox{\raise 2pt \hbox{$<$} \kern -13pt
                     \lower 3pt \hbox{$\sim$}}\ }
\def\greatersim{\ \hbox{\raise 2pt \hbox{$>$} \kern -13pt
                     \lower 3pt \hbox{$\sim$}}\ }

\def\lsim{\mathrel{\rlap{\lower4pt\hbox{\hskip1pt$\sim$}}
    \raise1pt\hbox{$<$}}}                % less than or approx. symbol
\def\gsim{\mathrel{\rlap{\lower4pt\hbox{\hskip1pt$\sim$}}
    \raise1pt\hbox{$>$}}}                % greater than or approx. symbol

\input epsf.tex
\def\desepsf(#1 width #2){\epsfxsize=#2 \epsfbox{#1}}

\newenvironment{tolerant}[1]{\par\tolerance=#1\relax}{ \par }
\usepackage{amsmath,bm}
\usepackage{lineno}
\usepackage{cite,mcite}
\usepackage{tikz}
\usepackage[symbol]{footmisc}

\providecommand{\DOI}[1]{\href{http://dx.doi.org/#1}}

\usepackage{wrapfig}
\usepackage{orcidlink}

\usepackage{xcolor}
\hypersetup{
     colorlinks  = true,
     linkcolor   = blue,
     anchorcolor = BrickRed,
     citecolor   = blue,
     filecolor   = blue,
     menucolor   = blue,
     runcolor    = cyan,
     urlcolor    = blue}

\begin{document}

\date{}

\title{
Science for Peace and the need for Civil Clauses\\ at universities and civilian research institutions
}
\author[1]{
J.~Altmann,
U.~Amaldi~\orcidlink{0000-0002-3425-8937},
M.~Barone~\orcidlink{0000-0002-2115-4055},
A.~Bassalat~\orcidlink{0000-0002-0129-1423},
M.~Bona~\orcidlink{0009-0003-7058-7927},
J.~Beullens,
H.~Brand~\orcidlink{0009-0008-0543-4875}, 
S.~Brentjes~\orcidlink{0000-0002-8205-8550}, 
D.~Britzger~\orcidlink{0000-0002-9246-7366},
J.~Ellis~\orcidlink{0000-0002-7399-0813},
S.~Franchoo~\orcidlink{0000-0001-7520-5922},
A.~Giammanco~\orcidlink{ 0000-0001-9640-8294},
A.~Glazov~\orcidlink{0000-0002-8553-7338},
C.~Heck~\orcidlink{0009-0005-1313-1661},
H.~Jung\thanks{Corresponding author and contact person: hannesjung@science4peace.com}~\orcidlink{0000-0002-2964-9845},
S.~Kraml~\orcidlink{0000-0002-2613-7000},
L.~L\"onnblad~\orcidlink{0000-0003-1269-1649},
M.~Mangano~\orcidlink{0000-0002-0886-3789},
M.~Renneberg,
Th.~Riebe,
A.~Sabio-Vera~\orcidlink{0000-0003-0228-5313},
R.~Sanders,
J.~Scheffran~\orcidlink{0000-0002-7171-3062}, 
M.~Schmelling~\orcidlink{0000-0003-3305-0576},
T.~Schucker~\orcidlink{0000-0001-5551-0444},
T.~Suzuki~\orcidlink{0009-0001-7710-8429},
A.~Tanasijczuk\orcidlink{0009-0008-7445-8002},
I.~Tsakov,
D.~Valls-Gabaud,
M.~Walker
}
\begin{titlepage} 
\maketitle
\vspace*{-9cm}
\begin{flushright}
\today
\end{flushright}
\vspace*{+6cm}

\centerline{The Science4Peace Forum\footnote[2]{Where authors are specified, they take responsibility for their respective sections. Their opinions are not meant to represent the position of the institution to which they are affiliated. When affiliations are mentioned, this is simply to illustrate the scientific contexts in which the authors work.}}

\begin{abstract}
After the end of World War II, the commitment to confine scientific activities in universities and research institutions to peaceful and civilian purposes has entered, in the form of {\it Civil Clauses}, the charters of many research institutions and universities. In the wake of recent world events, the relevance and scope of such Civil Clauses has been questioned in reports issued by some governments and by the EU Commission, a development that opens the door to a possible blurring of the distinction between peaceful and military research. 

This paper documents the reflections stimulated by a panel discussion on this issue recently organized by the Science4Peace Forum. We review the adoptions of Civil Clauses in research organizations and institutions in various countries, present evidence of the challenges that are emerging to such Civil Clauses, and collect arguments in favour of maintaining the purely civilian and peaceful focus of public (non-military) research.

\end{abstract} 
\end{titlepage}

\tableofcontents

\newpage

\section{Preface}

The Convention of the international organization CERN in Geneva states that ``the Organization shall have no concern with work for military requirements". Likewise, the international organization JINR in Dubna declares that its research results ``can be used only for peaceful purposes for the benefit of the whole of mankind". In Germany ``Civil Clauses", commitments to work only for civilian and peaceful goals, have been adopted as voluntary declarations by many institutions and universities.
However, whilst there has been common agreement on research for peaceful and civil purposes after World War II, recently discussion of the utility and relevance of Civil Clauses has been put on the table by the EU commission as well as by some national governments.

This development reflects new, concerning  changes in science policy. On one hand, following the invasion of Ukraine by the army of the Russian Federation, co-operation with scientists from Russian and Belorussian institutions was stopped and their scientists were excluded from participating in international civilian research projects. On the other hand, attempts to open for military purposes research institutions that were in the past bastions of civilian research contribute further to sanctioning their science communities and treating them as enemies, abandoning the universal nature of science.

Triggered in 2024 by a {\it White Paper} of the EU Commission as well as a {\it Position Paper} by the German Ministry for Education and Research, the Science4Peace Forum started in September 2024 discussions among a panel of experts on different aspects of Civil Clauses. In Spring 2025, the EU commission has released the ReArm Europe Plan/Readiness 2030.

In response to these developments, 
in this paper we collect arguments for maintaining the purely civilian and peaceful focus of public (non-military) research and argue that scientific progress for the benefit of humanity can only be achieved by the collective efforts of all countries and nations. Limiting the collaboration on research only to those countries that share the same political values will create only anger, mistrust and further conflicts. This will lead to  another arms race and will work against finding solutions to the most important problems humanity is facing now: climate change, ecocide, poverty and, most of all, the proliferation of wars.

\section{Introduction}
In January 2024, the EU Commission issued a {\it White Paper} \cite{EU-WhitePaper} calling for a special effort to promote research with both civil and military objectives (dual-use research).  Similarly, the German Ministry of Education and Research (BMBF) released a {\it Position Paper}~\cite{BMBF-2024} in March 2024 calling for the deepening of cooperation between civilian and military research institutions and for establishing {\it funding incentives for increased cooperation between civilian and military research}. In its annual report for 2024, the German {\it Research and Innovation Expert Commission} proposed dissolving the previous separation between civil and military research~\cite{efi-2024}. In March 2025 the EU commission has released the {\it ReArm Europe Plan/Readiness 2030}~\cite{Rearm2030}.

This reorientation is in fundamental contradiction to the spirit of civilian research following the experiences of World War II: 
at the opening of the Technical University of Berlin in 1949 Major-General E.P. Nares said in his opening speech~\cite{TU-Berlin-1946}: {\it The British authorities are well aware that the Technical High School of Berlin made a valuable contribution to your country's war potential and was one of the props of the technical development of the vast war machine which Hitler built up to oppress other peoples ... The four great Allies - Russia, France, the United States and Great Britain - have vowed that such a war machine shall never be allowed to rise again....} and he continued by saying {\it Science and technology can be and must be devoted to advancing the peace and civilisation of man and this can only be so if they are used with responsibility.}
The Science Council of Japan vowed in 1950  {\it to never become engaged in scientific research for war purposes}~\cite{civil-clause-Japan}. At the international organization CERN, where the Higgs Boson was found in 2012, the Convention of 1954~\cite{CERN-convention} demands explicitly that it {\it shall have no concern with work for military requirements}, and the guiding principles of the Helmholtz Research Center DESY in Germany~\cite{DESY-guiding_principles} stipulate that its {\it research pursues goals that are peaceful and serve civil society}. Moreover, many universities and institutions in Germany have adopted  Civil Clauses~\cite{Zivilklausel-frankfurter-Erklaerung} that voluntarily focus their research and teaching for purely civilian and peaceful purposes. 

The discussion of opening civilian research  institutions to military research threatens international scientific cooperation,  and  research funds that  are allocated to civilian research might be withdrawn and could be dedicated to dual-use and military research, as criticized in a statement~\cite{HRK-2024} of the German University Rectors' Conference (HRK). However,  in the ReArm Europe Plan/Readiness 2030 White Paper of the EU Commission~\cite{Rearm2030} it is explicitly stated that the {\it European Innovation Council and the planned TechEU Scale-up Fund will invest in dual-use technologies}.   

At DESY the directorate initiated a discussion  in June 2024 \cite{MoPo-DESY} on whether the restriction of research to civilian and peaceful purposes is still appropriate, or whether military research should be allowed at the laboratory. This initiative triggered many discussions and protests at DESY, where shortly after this announcement a petition in opposition to this move~\cite{DESY-petition-military-research} was launched. This topic has caught the attention of the national~\cite{Zeitenwende_der_spiegel_2025_1} and international press~\cite{Gibney_2025}. \\

We are facing significant changes in science policy:
\begin{itemize}
\item {\bf Sanctions in Science}  

Immediately after the start of the invasion of Ukraine by troops of the Russian Federation, several science institutions in Europe initiated sanctions against scientists from Russian and Belorussian Institutions. At DESY, the imposed sanctions~\cite{DESY-sanctions2} include a ban of common scientific publications and a ban for DESY scientists to participate in  scientific conferences attended by scientists affiliated with sanctioned institutions. The Science4Peace Forum has collected many arguments against excluding scientists from international cooperations~\cite{Albrecht:2023xxi}.  At CERN, collaborations were put on hold. At the end of 2023 CERN Council decided not to renew cooperation agreements with Russian and Belorussian institutes after their expiry dates in 2024. The Science4Peace Forum has warned about the long-term consequences of such a decision~\cite{Ali:2024fxg}. At least the cooperation of CERN with JINR~\cite{JINR}, which is also an international organization, was not cancelled, and is continuing under restrictive conditions 

These steps marked a clear change in science policy. Scientists from certain countries are at risk of being excluded from scientific cooperation, a step that did not happen previously at CERN even during  the Cold War (the only exception being in 1993 when Serbia and Montenegro were excluded from CERN  when it implemented a mandatory decision of the UN Security Council~\cite{cern-embargo-serbia1993,Abbott1993} aimed to stopping the fighting in (former) Yugoslavia).
\item {\bf Role of science in geopolitical strategies} 

In a report from the German National Academy of Science and Engineering (acatech) {\it Security, Resilience and Sustainability}~\cite{actech-2022} from 2022, an increased role of science in security policy and military research was proposed, but with the caution that: {\it  [...] if Germany positions itself too assertively [in increased spending for military], its current reputation as a peaceful nation could suffer.} In a report of the Leopoldina and Deutsche Forschungs Gemeinschaft (DFG) from 2024~\cite{leopoldina-2024-eng} it was argued  that: {\it [...]  as a result, science and innovation are increasingly being identified as geopolitical levers of power in Europe and North America in order to strengthen resistance and competitiveness in the interests of national security.}. Even fellowships and student exchange are seen critically: {\it  [...] no longer admitting young Chinese academics with a scholarship from the Chinese Scholarship Council (CSC) in the future }, as written in Ref.~\cite{leopoldina-2024-eng}.
\item {\bf Civil Clauses versus opening civilian research facilities for military research}  

The discussion about Civil Clauses and the focus of civilian research institutions on purely civilian research has gained momentum recently with a White Paper~\cite{EU-WhitePaper} of the EU commission, where it is complained that resources spent for purely civilian research are missing from funding for military (or dual-use) research. 
In the Spring of 2025 the EU Commission has released the {\it ReArm Europe Plan/Readiness 2030} report~\cite{Rearm2030}, explicitly emphasizing dual-use research and the contribution from AI, quantum computing  and cyber warfare not only for defensive but also offensive purposes.
With the further discussion on increased spending for military - at the moment several EU countries have already increased their spending to 2\% of GDP - and the request to increase the spending from 2\% to perhaps 3.5 or even 5 \%, the resources for pure civilian research will decrease, and institutions may be forced to look for additional funding, for example by removing Civil Clauses and opening their research facilities for military or dual-use research. While such an attempt might solve the short-term funding problem, it will lead to dramatic and significant changes in the nature of science and research.
\end{itemize}

In this paper we discuss these three topics in more detail, with the emphasis that science and especially fundamental science should be universal and independent of any political and geo-political strategies. We emphasize that sanctions and restrictions in science are counter-productive and lead only to further separation and confrontation, rather than helping solving international conflicts. We argue that science should play a positive role in international affairs,  so that the language of science is used to build bridges and dialogue, instead of banning communication. We also argue that successful science serving society and helping solve the most important problems of humanity must be international, including all countries. The results of this science must be accessible to the whole world openly and freely, something that can be guaranteed only by purely civilian projects.

\section{Sanctions in Science and the role of geopolitics}
Now is the first occasion  that sanctions are imposed against scientists in national and international institutions without the mandate of the UN Security Council. The ban from CERN on scientists from Russian and Belorussian institutions by the end of 2024 (when the umbrella international co-operation agreements ended) has had many negative consequences, ranging from the loss of resources, knowledge and know-how, to, likely more importantly, the deep frustration of the scientists who were discriminated against and may now be motivated to engage in military research. 
 It will take a very long time until this damage in scientific cooperation can be healed, and many scientists who were expelled say that, even after the sanctions  are lifted at CERN, they would refuse to return.

Not only do sanctions in science damage scientific cooperation and question the universality of science as well as hinder necessary development, for example in climate research where cooperation is essential to access necessary data, also the acceptance of sanctions by many in the scientific community marks  a major step in handing over independence to political leaders. This step cannot be overemphasized, as German history in particular  shows clearly the dangers when scientists work in the name of politics, without feeling responsibility for the consequences of their research, leaving that to politicians.

Science always depends on the political framework, which impacts how  public funding is distributed across different programs. 
Whilst choosing a funding source  remains a responsibility of the institutions and of the individual scientist, this choice may have consequences that go beyond their control. For example, if the political situation changes during a funding period, and politics is imposing new constraints on  scientific projects that have already been funded, a direct intervention of politics into science may occur. This shows that, even during a funding period, the leaders of institutes might experience political pressure, that they are unable to resist, emphasizing the  individual responsibility of every scientist. 

In the report of the Leopoldina and DFG ~\cite{leopoldina-2024-eng} concerning international cooperations it was explicitly argued  that {\it there is growing political pressure to stop cooperating with these countries, or only cooperate to a limited extent, in order to secure strategic competitive advantages and one's own research integrity, avoid dependencies and not indirectly support developments abroad that are questionable from a Western perspective.} According to these statements, science's goals should be clearly dominated by political reasoning. These considerations imply that sanctions against scientists are part of a general strategy, not  motivated so much by specific political reasons and not directly related to Russia's war against Ukraine, since these strategies were  already in place.

The clear mention of the {\it Western perspective} bears the risk of decoupling the research and development of western countries from the rest of the world, a development which is  counterproductive given the increasing influence and potential that non-western countries have in new research advances (e.g., the development of DeepSeek~\cite{DeepSeek2025}).

It is further argued in Ref.~\cite{leopoldina-2024-eng}  that {\it academic research also bears responsibility for safeguarding our basic democratic order and other national values and that it can no longer be pursued solely for its own sake.} If such statements indicate the direction for future scientific research and development, the universal and humanistic value of science is put into question, and science would lose all its objectivity and credibility, being degraded to a tool for enforcing political strategies.

\section{Civil Clauses: past and present}
The concept of Civil Clause in Germany emerged from the experience of the connection of science and scientists with the German military and its consequences in World War II, as a commitment to perform research only for civil, non-military and peaceful purposes. Civil Clauses go back to the foundation of the Technical University of Berlin~\cite{TU-Berlin-1946} in 1946.  An overview of existing  Civil Clauses in Germany is given in Ref.~\cite{Zivilklauseln-to-date}, a writeup of a lecture series on Civil Clauses  is given in Ref.~\cite{2012:nielebock:zivilklaus}. Beyond Germany, the Science Council of Japan declared in 1950, {\it its commitment to never become engaged in scientific research for military purposes}~ \cite{civil-clause-Japan}.

\subsection{Some historical remarks on science in Germany \\
Author: Mark Walker, John Bigelow Professor of History,
Union College, Schenectady, NY USA, Monika Renneberg}

German scientists had a strong and consequential relationship with militarism in the first half of the twentieth century. During the First World War, most German scientists served as regular soldiers, but a significant minority put their professional expertise to work on early radar systems, aerodynamics for aircraft development, and of course chemical weapons. The overwhelming majority of these scientists supported the German war effort uncritically. Here Albert Einstein was the exception who proved the rule. When Germany was defeated, most scientists, like most Germans, focussed more on the harsh terms of the Treaty of Versailles than what responsibility Germany shared for the war.

When the Nazis came to power in 1933, they began massive investments in rearmament. It is important to note, however, that during this period Hitler publicly insisted that he wanted peace, not war, and the rearmament was only so that Germany could protect itself. When the Second World War began with the invasion of Poland, most scientists, including those opposed to Nazism, were either drafted into the armed forces or found research and development projects that allowed them to serve the war effort as scientists. These projects were similar to those in other countries, like mines, submarines, radar, rockets, aeronautics, and even nuclear weapons. When the war turned against Germany after the defeat at Stalingrad and more and more German men were drafted, scientists came under even more pressure to find positions considered “indispensable for the war effort.” When the war was over, the four victorious powers invited or sometimes kidnapped those German scientists whose research was of military interest. Here the most prominent example was Wernher von Braun and long-range ballistic missiles (for references and further reading see~\cite{Renneberg1994,Heim2009,Walker2024}).

\subsection{Commitment not to engage in research for military purposes - the case of Japan \\
Author:   Tatsujiro Suzuki
Visiting Professor, Nagasaki University, and President, Peace Depot, Japan}

After the end of World War II, the scientific community of Japan established the Science Council of Japan (SCJ) in 1949, set up as an {\it independent and special} organization under the jurisdiction of the Prime Minister of Japan. In 1950,  the SCJ adopted a statement on its commitment never to become engaged in scientific research for war purposes. In 1967, the SCJ issued again a similar statement not to engage in research for military purposes. 

The spirit of those pledges has been maintained until now, but the worsening security environment has pushed the government to encourage scientists to work for so-called {\it dual-purpose}  research, whose purposes could include military use.

 In 2015, the government established a new research fund, called the {\it National Security Technology Research Promotion}  sponsored by the Acquisition, Technology and Logistics Agency (ATLA) of the Ministry of Defense (MOD)~\cite{ATLA-Japan}. In this funding scheme, research proposals are invited and reviewed with a clear objective of supporting projects which are likely to be applied for development of defense (military) equipment or systems. This is the first time that the Ministry of Defense directly funds university research projects. 
 
This funding scheme caused a major academic debate in Japan. Responding to this debate, the SCJ issued a report~\cite{Report-Japan-2017} and a statement on research for military/security research in 2017~\cite{civil-clause-Japan}. In that statement, the SCJ reaffirmed the previous two statements on  its commitment not to engage in research for military purposes. The statement says the following:
{\it It should be pointed out that this funding program (of the MOD) has many problems due to these governmental interventions into research. From the standpoint of a sound development of the sciences, funding should be increased further for research in civilian areas where autonomous research by scientists and unrestricted publication of research results are assured.}

\subsection{The Civil Clause in Germany \\
Author: Jonathan Beullens}

Two and a half months after the liberation of Germany from fascism on May 8th 1945 by Allied forces (whose 80th anniversary is taking place this year), a framework for the future of the German people was agreed upon at the Potsdam Conference. This included the famous four D's: Denazification, Democratization, Decentralization and Demilitarization. As German schools and universities had played an essential part in the ideological indoctrination of children, the development of weapons systems as well as pseudoscientific theories in favor of conquering  {\it Lebensraum}~\cite{Heinemann2021}, the Potsdam Agreement contained the following passage in its {\it Political principles}:
{\it German education shall be so controlled as completely to eliminate Nazi and militarist doctrines and to make possible the successful development of democratic ideas.}~\cite{PotsdamAgreement}

When the Technical University of Berlin (TUB) was to start up its activities again in 1946, the conscious decision was made not to label it a {\it reopening}, in order to make a decisive break with its Nazi past~\cite{TU-Geschichte}. A humanities faculty was founded to promote societal responsibility in science and the reflection of (natural and technical) scientists on the consequences of their work. To manifest this change in the way science should be performed, the institution was renamed: the  {\it Technical High School of Charlottenburg} should henceforth be called the Technical University of Berlin. In a speech at the opening ceremony~\cite{TU-Berlin-1946} the  British Major-General E.P. Nares stresses the expectations that would now govern the university:
{\it The implications of this change of name are simple but of vital importance. It should teach you that all education, technical, humanistic, or what you will, is universal: that is to say it must embrace the whole of man, the whole personality, and its first aim is to produce a whole human being, capable of taking his place responsibly beside his fellows in a community. Its second aim may be to produce a good philologist, a good architect, a good musician or a good engineer.} 

{\it But if education does not assist the development of the whole personality it fails in its aim, and this Technical University must not fail in its aim. You cannot bring into this building only the technical part of your minds and leave the other parts of your personalities outside or hang them up with your hat and coat on a peg in the hall. You must bring to your work all that you have - your love of art, your religion, your philosophy of life as well as your technical capacities - and allow them to develop together with your work through your experience here and your contact with your teachers and fellow-students. [...] }

{\it This universality is necessary in education because only by cultivating the whole of himself can man acquire a sense of responsibility, and only by responsibility can freedom, peace and justice - that is the happiness of all men - be assured.}

One of the first measures taken by the occupying forces in higher education in fulfilment of the principles of social responsibility and peace was the implementation of a ban against military research at TUB, which had participated in the research program for the development of the V2 rockets used to bombard London and Antwerp~\cite{Neufeld1995}. So the first Civil Clause in Germany was born as a direct consequence of fascist and militarist history. This Civil Clause was reaffirmed in 1991 by the senate of TUB, {\it out of responsibility and because of the university's role before and during the Second World War, especially in armaments research}~\cite{ZivilklauselTUBerlin}.

Ten years later, in 1956, the second Civil Clause was implemented at the newly-built nuclear research centre in Karlsruhe. With the luckily failed Nazi {\it Uranprojekt} to build nuclear weapons still fresh in the minds of the Allies, pressure was applied on the Adenauer administration that the research center should only follow peaceful objectives.

Broader efforts for demilitarization shaped the development of the new German constitution, which was passed in 1949. In its preamble it is stated that Germany should become "an equal partner in a united Europe" to "promote world peace" and its very first article declares the "inviolable and inalienable human rights" to be the basis for international coexistence. This includes article 26 of the Universal Declaration of Human Rights~\cite{UNHumanRightsDeclaration} concerning education, which asserts that 
{\it education shall be directed to the full development of the human personality and to the strengthening of respect for human rights and fundamental freedoms. It shall promote understanding, tolerance and friendship among all nations, racial or religious groups, and shall further promote the activities of the United Nations for the maintenance of peace.}

The preamble of the constitution of the United Nations Educational, Scientific and Cultural Organization (UNESCO)~\cite{UNESCO-constitution} affirms this goal, declaring
{\it that since wars begin in the minds of men, it is in the minds of men that the defenses of peace must be constructed;
That ignorance of each other's ways and lives has been a common cause, throughout the history of mankind, of that suspicion and mistrust between the peoples of the world through which their differences have all too often broken into war;
That the wide diffusion of culture, and the education of humanity for justice and liberty and peace are indispensable to the dignity of man and constitute a sacred duty which all the nations must fulfil in a spirit of mutual assistance and concern; [...]}

{\it In consequence whereof [the parties to this constitution] do hereby create [UNESCO] for the purpose of advancing, through the educational and scientific and cultural relations of the peoples of the world, the objectives of international peace and of the common welfare of mankind for which the United Nations Organization was established and which its Charter proclaims.}

Thus, Civil Clauses are to be seen as the fulfillment of the fundamental principles underlying worldwide social, economic, ecological and cultural development. Their implementation in scientific institutions contributes to the building of mutual trust between nations. The possibilities of open dialog created by focusing on civil research are essential to promoting peace and amity between peoples: Those who talk with each other are less likely to shoot at each other. In times of increasing hostility and wariness between nation states, especially concerning scientific cooperation, Civil Clauses provide generalizable criteria for international collaboration of German institutes not only with China but also with Turkey, Iran, Israel, the United States and other countries.

They are also a direct expression of the will for peace of the university members themselves: in Germany, the majority of Civil Clauses have been instituted after lengthy discussions between researchers, students and administrative staff in committees and senates, in part even by popular vote among the student bodies, as was the case, e.g., at the University of Frankfurt. At the beginning of the previous decade student protests against the introduction of extra tuition fees as well as calls for the improvement of working and learning conditions led to a discussion regarding the goal of research. The consensus emerged that universities and science cannot be truly free if financial constraints are in place that could force their opening up for military funded research. Freedom of science should be determined not only through the absence of state interference but also through the positive concept of enabling emancipatory and peace-minded work. Only in conjunction with the fulfillment of human dignity (Art. 1 of the German Constitution) and the principle of a social state order (Art. 20) can freedom of science (Art. 5) truly flourish and be of use to all humankind.

This stands in stark contrast to current attempts by state and federal governments to undermine Civil Clauses. These top-down approaches are, in light of the historical developments outlined above, in open disregard for lessons learned from two world wars as well as the social and ecological challenges of our time. Geraldine Rauch, mathematician and rector of the Technical University of Berlin, summarized succinctly what is necessary instead~\cite{Rauch2022}:
{\it The role of universities is not to settle military and political conflicts, but to carry out research and teaching in the interests of a more stable, social and sustainable world - this brings us all real security}.

\subsection{Civil Clauses in the Academy of Arts and Media  \\
Author: Christoph Heck, Academy of Media Arts Cologne (KHM) }

New media and technologies are increasingly shaping our everyday lives and warfare. Thus, their designers also bear an increasing social responsibility, which also poses new challenges for their educational institutions.
The role of media, art and cultural workers has changed in recent years due to the rise of social media, which has created a breeding ground for polarization, disinformation and propaganda in the information and media war. The same media we use to connect, to inform and to communicate with each other are being used for military operations and to construct AI-supported cyber-weapons and targeting and decision support systems~\cite{ZivilklauselAtKHM}.
As academies of art and media  are places of creation and reflection, they teach and conduct research at the intersection of aesthetics, technology, and society. They play an active role in the debate on cultural participation, civil values, creative power and social responsibility in the digital age.
Therefore, for an academy of  art and media, anchoring a Civil Clause in its basic regulations, which explicitly prioritizes {\it the peace-building and peace-preserving aspects of media in studies, teaching and research},  i.e.,  promoting the social aspects within social media and rejecting any involvement of art, science and research that serves or is aimed at waging war, is a clear position to adopt~\cite{Zivilklausel-KHM}. Since 2024, the Academy of Media Arts Cologne (KHM) has been committed to exclusively peaceful and non-military goals through a Civil Clause. This is particularly important at a time when societal rearmament and preparation for war is increasingly being socialised through the media.

\subsection{German Science Institutions and Civil Clauses \\
Author: Sonja Brentjes}

All major German research institutions  are financed by the federal government and various of its ministries. The ministry responsible for education and research (changing its precise designation depending on decisions of the respective government), for instance, declared in 2020 that it supports exclusively civil research \footnote[3]{This commitment has changed now with the new Coalition Agreement~\cite{Koalitonsvertrag-2025}.}. The defense ministry, in contrast, has since the mid-1950s financed military research at partly civil research institutions such as the Fraunhofer Society. The vulnerability of the research institutions to shifts in the political orientation of a government has become very clear during recent years, when debates about Civil Clauses and military research have intensified. In 2014, the main German research institutions, the German Science Foundation (DFG), and the Leopoldina, the National Academy of Sciences, decided to create a  "Joint Committee on the Handling of Security-Relevant Research"~\cite{Leopoldina-2018_2,ProgressReport2022}. The foundational document reveals that the inspiration for this step came from the federal government in 2012. 
However, since 2022 the activity report has been arguing for the need to react to geopolitical changes~\cite{Leopoldina-dualUse-2022}. The shift in political rhetoric and orientation of the federal government is reflected explicitly in the preface written by the presidents of the two institutions to the activity report of November 2022~\cite{ProgressReport2022}. This new orientation is expressed even more clearly in the title of the report of 2024. Whereas all previous reports were simply labeled Progress Reports, the 2024 report published in March 2025, carries now the title {\it Scientific Freedom and Security Interests in Times of Geopolitical Polarisation}~\cite{leopoldina-2024-eng}. This development raises serious questions about the stability of the peaceful orientation of research at universities and research institutes represented through these two institutions and their partners -- the Max Planck Society, the Helmholtz Society, the Fraunhofer Society and the Leibniz Society -- and therewith the roles of Civil Clauses at those institutions that adopted them in various forms.

The last four activity reports of the Joint Committee have primarily focused on issues arising from dual-use problems in the sciences and technology, in particular IT, AI, and biotechnology, as well as medicine and psychology. Civil Clauses were discussed as tools to restrict military research. This focus of Civil Clauses was defined as insufficient for the broader dual-use issues that the Joint Committee means to address. A second problem with Civil Clauses seen by the Joint Committee on the basis of a paper by a junior legal scholar concerns their relationship to the German constitution. Depending on their specific formulations and their legal status as either a part of a law regulating the universities and comparable teaching and research institutions or as a voluntary declaration of a university, they may contradict Art. 5.3 of the constitution regarding the freedom of research.

The interesting aspect of the declarations and reports of the Joint Committee concerns the insistence on academic freedom being consistent with self-regulatory activities regarding dual-use issues. On the institutional level, such activities should be implemented by creating ethics commissions at each institution, whose members should counsel researchers and research institutions about how to deal with dual use issues. Since all reports of the Joint Committee continue to emphasize freedom of research and teaching and the obligation to undertake such activities for the sake of improving human conditions, fostering peace, defending human rights, and protecting the environment, multiple spaces remain for integrating the existent Civil Clauses and their practical implementations into the sponsored development of responsible research and teaching with regard to dual-use issues. It even seems to be  possible to formulate new Civil Clauses in this broader framework. Such opportunities are  strengthened by the explicit condemnation of research for developing sanctioned weapons such as biological weapons or of research projects in explicit collaboration with armament industries, with the caveat that military research is acceptable if  the weapons are specialized to clean up the debris of wars such as land mines.

However, the increasingly confrontational language in the reports of the Joint Committee sends warning signals calling for a public debate of all aspects of dual-use issues beyond the limited circles of the meetings organized by the Joint Committee.

\subsection{Civilian and military research in different countries}
In the following we provide an overview on relation of military research at civil and public universities in different countries.

\begin{itemize}

\item{\bf Austria} \\
The Boku (Universit\"at f\"ur Bodenkultur, Vienna) has an Ethics Charter with an explicit commitment to civil society~\cite{Austria_1}. This seems to be the only university in Austria which has such a clause.  The Austrian Academy of Science, OeAW, on the other hand, has a commission for scientific cooperation with departments of the Federal Ministry of Defense (Bundesministerium f\"ur Landesverteidigung, BMLV). This commission, called MILKOM, supports the {\it fulfillment of all scientific requirements for comprehensive national defense in the mutual and national interest}~\cite{Austria_2}. The objectives as stated in Ref.~\cite{Austria_3}  are i) to provide material and financial support for basic research at the Academy on issues that also have a certain relevance to the BMLV and ii) to give the BMLV the opportunity to assign unmet research needs within its own area to members of the OeAW.

\item {\bf Belgium} \\
In Belgium, Civil Clauses have evolved significantly over the past decades. Belgium’s education and research systems are largely divided along linguistic lines, with each language community (Dutch-speaking and French-speaking) managing its own institutions. The three Regions (Flanders, Wallonia and Brussels) also contribute to research funding. Only a few federal bodies operate across these divisions. As a consequence, scientific policies in the two halves of the country sometimes differ, following different evolutions of the public opinion in the two main cultural communities.

Starting in the 1990s, Flanders maintained strongly pacifist regulations that effectively prohibited government funding for military research at civil institutions. This changed in 2018, when Minister Muyters introduced new guidelines allowing public funding for military research, provided it did not support offensive weapons development~\cite{Belgium-a,Belgium-b,Belgium-c}. These were broad policy changes, not specifically targeted at universities. More recently, the regulations have been further relaxed, making public funding for military-related research even more broadly available. However, the shift has sparked ongoing debates within various universities about the ethical implications and boundaries of such funding~\cite{Belgium-d}.

In Wallonia, military research is actively supported through both regional and federal initiatives. The region hosts major defense manufacturers such as FN Herstal and John Cockerill Defence, and the Belgian Ministry of Defence has announced a €1.8 billion investment in research and development under the STAR (Security, Technology, Ambition, Resilience) plan~\cite{Belgium-e}.

As a response to the war in Gaza, and prompted by the mobilization of Belgian students, the Rectors’ Councils of the two linguistic communities of Belgium (VLIR for Dutch-speaking and CREF for French-speaking universities) issued joint statements stating that they would have reviewed their ties with Israeli institutions~\cite{Belgium-f}, also referring to the principles previously outlined in the joint VLIR human rights assessment of 2019~\cite{Belgium-f2}, and in May 2025 they also called on the European Union to suspend its association agreement with Israel~\cite{Belgium-g}. As an example, the Université catholique de Louvain, in Louvain-la-Neuve, has enforced since December 2024 an Ethical Charter for Responsible Partnership~\cite{Belgium-h}. In practice, all new and ongoing international collaborations (including EU-funded networks) are reviewed, and those that have potential for dual use and that include partners in countries deemed problematic from the point of view of respecting human rights or international law, are reportedly interrupted or vetoed. Currently, the list of problematic countries includes Iran, Israel and Russia.

\item {\bf France} \\
In some cases there is a rather strict separation between civilian and military research in France, in others not. For nuclear physics and applications, the Commissariat {\`a} l'{\'e}nergie atomique et aux energies alternatives (CEA) maintains a separate Direction of Military Applications (CEA/DAM) that is responsible for nuclear-weapon research and maintenance. More research that is explicitly military takes place within the defense-related companies themselves~\cite{Franchoo-Frace-2025}. 

The Centre National de la Recherche Scientifique (CNRS) has as its mission to carry out {\it all research of interest to the advancement of science as well as to the economic, social and cultural progress of the country} and {\it to contribute to the application and promotion of the results of this research} (Article R322-2 of the Code de la Recherche Ref.~\cite{France_CNRS}). While this implies mostly civilian research, dual-use or even military applications are not excluded.

An interesting element is the introduction by the French senate in 2022 of the obligation for newly-qualified doctors to take an oath that focuses on scientific integrity in their future career~\cite{Doctorate-Declaration-2022}.

\item {\bf Germany } \\
Civil Clauses exist in more than 70 universities and research institutions~\cite{Zivilklauseln-to-date}. Civilian and military research is separated, military research is performed in special research institutes such as the  Fraunhofer Society~\cite{Fraunhofer-Militar}, while civilian research is performed at universities and research centers like DESY~\cite{DESY-guiding_principles}, GSI~\cite{GSI-Vertrag} or KIT~\cite{Zivilklausel-KIT}.
Despite this, the federal Bavarian state has approved a law in 2024 which pushes universities to cooperate with the army and explicitly excludes any Civil Clause~\cite{Bundeswehr-Bayern}.

\item {\bf Greece}\\
 In Greece, researchers also have the freedom to choose their field, but there is no clear separation between civilian and military research, and researchers can receive funding for both~\cite{Barone-Greece2025}.
 
\item {\bf Israel}\\
In Israel military and civilian research is conducted at public universities, as shown by examples from Tel Aviv University~\cite{TelAviv} and the Technion~\cite{IAAC}.

\item {\bf Italy}\\
In Italy, there is a distinction between basic research and military research, which are carried out by different institutions. However, there is no Civil Clause in Italy~\cite{Barone-Italy2025}.

\item {\bf Japan}\\
The Science Council of Japan is committed to non-military research~\cite{civil-clause-Japan}.

\item {\bf Palestine} \\
The document on research topic priorities in Palestine mentions only purely civilian research

\item {\bf Sweden} \\
In Swedish universities there is nothing corresponding to a Civil Clause, and in general there is seldom any explicit distinction made between civilian and military defence research.

Recently there has been a strong focus in Sweden on defence and resilience in general, and a strategic co-operation with universities~\cite{Sweden-CampusDefense} was initiated to strengthen the defence sector in Sweden through education and research. This has resulted in a number of new defence-oriented education programs (see eg.~\cite{Lund-2025-Preparedness}). There is a very strong focus on civilian issues in this collaboration, but there is no mention of excluding military research, and the Swedish Defence University, which certainly engages in military research~\cite{Swedish-defence-university}, is a part of the collaboration.

From an overview of research related to civilian and military defence by the defence committee of the Swedish parliament~\cite{Swedish-Palrliament-2008} it is clear that the vast majority of research performed by the Universities is of civilian character, but there are some exceptions, most notably when it comes to the development of manned and unmanned aerial vehicles, where {\it dual-use} is emphasised.

\item {\bf United Kingdom}\\
The primary source of civilian research funding in the United Kingdom is UK Research and Innovation (UKRI)~\cite{UKRI}, an independent government body sponsored by the government's Department for Science, Innovation and Technology. Military research is funded through the UK Ministry of Defence. UKRI states as its vision {\it ... an outstanding research and innovation system in the UK that gives everyone the opportunity to contribute and to benefit, enriching lives locally, nationally and internationally} with the mission {\it to convene, catalyse and invest in close collaboration with others to build a thriving, inclusive research and innovation system that connects discovery to prosperity and public good.} It manages nine Research Councils active in specific areas: the Arts and Humanities Research Council (AHRC), the Biotechnology and Biological Sciences Research Council (BBSRC), the Engineering and Physical Sciences Research Council (EPSRC), the Economic and Social Research Council (ESRC), Innovate UK, the Medical Research Council (MRC), the Natural Environment Research Council (NERC), Research England and 
the Science and Technology Facilities Council (STFC). The latter is the primary UK funding agency for CERN and related research, and therefore responsible for ensuring that the UK scientists that it funds to collaborate with CERN respect its Convention.

\end{itemize}

\section{Science and Military: National and International Research Institutions}
\subsection{From Open Science to Military Secrecy: The Risks for DESY \\
Author: Alexander Glazov, DESY}

DESY plays a crucial role in international cooperation. Historically, projects such as the HERA e-p collider benefited from extensive collaboration with numerous European countries, including France and Italy, as well as Russia (initially the USSR), the United States, and Japan. Over the years, DESY scientists have actively participated in major international collaborations at CERN and KEK~\cite{kek}, engaging with researchers from across the globe. With the commissioning of PETRA III, one of the largest synchrotron light facilities in the world, DESY became a major international hub to study fundamental phenomena in condensed matter, plasmas and molecules, and on the structure and function of complex materials to biomolecules and cells~\cite{DESY-photon-science}. While DESY is a national laboratory, it is widely recognized as a key international player, regularly reviewed by international committees that consistently encourage collaboration.

Peaceful scientific research conducted by scientists is fundamental to the development of humanity as a whole. Common fundamental goals foster a collaborative environment among researchers from diverse cultural and political backgrounds. The exchange of ideas between different scientific schools is crucial for innovation, promoting both the generation of new concepts and mutual respect for collective achievements. Open scientific hubs such as CERN, KEK, and DESY play an essential role in nurturing these interactions, strengthening the global research community.

International research is also vital for fostering trust among scientists from different nations. Historically, physicists' opinions have played a significant role in shaping governmental perspectives and promoting trust between nations. The end of the Cold War would not have been possible without Soviet leaders trusting that there was no need for ideological conflict with the West and recognizing that both sides shared fundamentally common values. The phone call between M. Gorbachov and A. Sakharov while A. Sakharov was still in exile was one of the main turning points of perestroika. This phone call would not have been possible without the influence of Gorbachov's scientific advisors, who advocated for peaceful cooperation with the West based on Sakharov's concept of convergence.

Fundamental scientific research is inherently dependent on large-scale projects requiring substantial investment. Most cutting-edge facilities built in recent years have relied on contributions from participating nations. A prime example is the European XFEL, operated by DESY, where Russia contributed approximately 27\% of the construction costs (up to 300 million euros by 2017~\cite{xfel-russia}. The war in Ukraine has called into question the interconnectedness of financial interests. However, investments in fundamental research have little connection to military activities. The involvement of DESY in military research would jeopardize such investments, making future contributions from international partners highly unlikely. Thus, while short-term financial gains from military research might seem attractive, they could ultimately become a limiting factor for the laboratory's long-term sustainability and international credibility.

Fundamental physics research, and DESY in particular, have been leading forces in open-access research. Open access is a foundational principle of fundamental science and remains the prevailing model for publicly funded research in Europe. Open-source and open-access frameworks have significant industrial implications, as exemplified by the recent Chinese DeepSeek deep learning model~\cite{DeepSeek2025}, which could lead to considerable reductions in electricity costs and a smaller environmental footprint. Military research, however, fundamentally contradicts the principles of open access. European science, and DESY in particular, should maintain leadership in open science, serving as a global example of transparency and collaboration.

The introduction of military research into existing civilian-only facilities raises multiple security concerns. Open access to research infrastructure would necessarily be restricted, reducing opportunities for international collaboration. Additionally, laboratories involved in military research become attractive targets for cyber threats from adversarial entities and, in extreme cases, could be considered legitimate military targets during conflicts. The potential for such attacks is particularly alarming given DESY’s location in the heart of Hamburg. Any form of military-related targeting of DESY could have catastrophic implications for the city's civilian population. Moreover, DESY employees themselves would face greater security risks. While recent security measures, such as the introduction of two-factor authentication for DESY computing centers, make direct cyber intrusions more difficult, they also increase the likelihood of personal attacks aimed at obtaining authentication devices, placing additional risks on employees. The risks would multiply significantly if military research were introduced.

Furthermore, many physicists have chosen fundamental research precisely because of its strictly non-military nature. These researchers joined civilian institutions like DESY under the assumption that their work would remain uninvolved in military applications. If military research were introduced, they would face an ethical dilemma: continue working in an environment contrary to their beliefs or leave, despite having dedicated much of their careers to their research at DESY.

If military research must be expanded, it should be conducted in dedicated facilities located in less populated areas, designed with comprehensive security measures and clear contractual agreements that align with the nature of military-focused work. This approach would be far more appropriate than compromising the integrity and mission of existing civilian research institutions.

\subsection{Research at CERN: "no concern with work for military requirements" \\
Author: John Ellis, King's College London}

CERN was founded shortly the end of the Second World War with the explicit intention of bringing together the scientific communities of European countries that had previously been fighting each other, to collaborate, in the words of its Convention~\cite{CERN-convention}, on ``research of a pure scientific and fundamental character, and in research essentially related thereto". It is widely known that the CERN Convention goes on to state that the Organization "shall have no concern with work for military requirements". The French version of the CERN Convention states that the Organization ``s'abstient de toute activit{\'e} {\`a} fins militaires", which corresponds more clearly, perhaps, to ``refrains from any activity for military purposes". To my knowledge, this rule has always been strictly applied, to the extent that CERN has refused to collaborate with institutions that are involved in military research. ``Science for Peace" is in CERN's genes.

It is worth recalling that scientific projects hosted by CERN are mainly conducted by some 12000 external researchers from over 100 countries who are attracted by CERN?s programme of peaceful international research in fundamental science.  The majority of these external scientists are affiliated with institutes in its Member States, which have accepted the provision in the CERN Convention that it will refrain from military research. Other external scientists are affiliated with institutes in countries that have signed international co-operation agreements with CERN, which specify that they  {\it shall use the results of their co-operation for non-military purposes only}. Similar wording appears in the Memoranda of Understanding governing individual research projects such as experimental collaborations - see, e.g., \cite{TVLBAIMOU} - and the development of accelerators, including the FCC~\cite{FCCMOU}. Individual researchers working at CERN are considered to be members of its personnel and expected to act in accordance with CERN's Code of Conduct, including its peaceful mission. However, it is difficult to monitor or control that this  expectation is respected by all individual external scientists.

I once accompanied a CERN Director-General who shall remain nameless to on a visit to a research group that shall remain nameless in a country that shall also remain nameless, with the aim of expanding scientific collaboration. The discussion started swimmingly. Unfortunately, I had to rush out for a moment to attend to an urgent call of Nature. When I returned a few minutes later, the atmosphere had turned distinctly chilly, and the meeting broke up prematurely. I discovered later that our prospective partners had revealed they worked on military laser systems, and that put a stop to any discussion of collaboration. 

On another occasion, another nameless institute in a different nameless country approach- ed CERN about collaboration. However, due diligence revealed that the institute worked on ballistic missiles, so that was the end of that discussion.

In some cases, however, CERN does collaborate with research establishments that do some military research. However, these are institutes - both in CERN Member States and elsewhere - that have fences across their campuses separating groups working on military research from their civilian colleagues. Only the latter may collaborate with CERN.

DESY is one of CERN’s most valued research partners, and it is reasonable to ask what would happen to this partnership if DESY were to accommodate dual-use research. If the research with military applications were separated from the civilian researchers by a fence, the collaboration could perhaps continue unimpeded. But does DESY really want its campus to be divided by a fence? I fear that the character of the institute would change irreversibly for the worse.

After the statement against military research, the CERN Convention goes on to state that ``the results of its experimental and theoretical work shall be published or otherwise made generally available.” CERN takes this provision very seriously, insisting (like many funding agencies) that all its research be published under ``Open Access" rules. To my mind, this provision places another significant obstacle in the way of collaboration with institutes that perform dual-use or military research. CERN has recently extended the principle of openness to include Open Infrastructure, Open Source Software and Open Hardware.

The insistence on avoiding any association with military research has played a key role in enabling CERN to become a global research centre. As such, it has hosted research teams from many countries that are political or military adversaries, such as India and Pakistan, Israel and Palestine, Iran and the US, even until recently Russia and Ukraine. As the founders of CERN recognised, research into the fundamental laws of physics is a topic of universal interest and value, that advances best if it is open to all scientists whatever their origin and its results are published in the open scientific literature. Until now, DESY has adopted a similar policy of openness, and has also developed into a global research centre. It is to be hoped that DESY will be able to resist any political pressures to change this policy, and avoid harming its enviable reputation. 

\subsection{Civilan and military research in Spain \\
Author: Agustin Sabio-Vera} 
In the current geopolitical context, Spanish universities and public research institutions, such as the Consejo Superior de Investigaciones Cientificas (CSIC), have become strategic actors due to their capacity to generate knowledge with dual-use potential (scientific and technological outputs with both civilian and military applications). This is exemplified by CSIC's 2023 collaboration with the Spanish Air and Space Force under the BACSI initiative, aimed at jointly developing intelligent, sustainable, and connected military bases~\cite{Spain_1}. Similarly, entities such as ISDEFE (Ingenieria de Sistemas para la Defensa de Espana) have formalized a growing number of agreements with universities (e.g., the Universidad de Salamanca~\cite{Spain_2}) to support research in areas like energy efficiency, emergency technologies, and emerging defense-related innovations.

These partnerships reflect the increasingly blurred boundaries between civilian research and military objectives. While this integration brings opportunities for funding and technological advancement, it also raises serious ethical and legal questions about how publicly-funded scientific institutions manage the risks of dual-use research. This debate has intensified in recent months due to revelations that some Spanish universities participate in EU-funded research projects involving Israeli defense firms such as Israel Aerospace Industries (IAI). Investigative journalism has shown that research financed through the European Defence Fund and Horizon Europe has, in certain cases, ended up benefiting companies whose technologies are reportedly used in conflict zones like Gaza~\cite{Spain_3,Spain_4}. These programmes, while designed to strengthen innovation and competitiveness in the EU's security and defense sector~\cite{Spain_5,Spain_6}, have drawn criticism when funding flows indirectly to actors involved in armed conflicts.

These developments have triggered widespread student protests and civil society backlash, calling for a halt to collaborations with military-linked institutions and demanding transparency and ethical accountability from academic leadership. Against this backdrop, Regulation (EU) 2021/821~\cite{Spain_7}, directly applicable in Spain, requires institutions to control the export of sensitive technologies, including intangible transfers such as emails, cloud-based data, or international publications. Although it includes exemptions for `basic research' and public domain knowledge, the regulation's vague definitions leave universities vulnerable to legal uncertainty and ethical ambiguity. To mitigate these risks, the regulation promotes the adoption of Internal Compliance Programs (ICP), now increasingly expected within Spanish research institutions. The Spanish government, through the Subdireccion General de Comercio Internacional de Material de Defensa y de Doble Uso, issues licenses and guidance~\cite{Spain_8}. The CSIC has begun establishing internal compliance structures, and similar measures are being encouraged in other institutions engaged in defense-related research.

In this evolving and politically sensitive landscape, there is an urgent need for clearer national legislation to regulate how public science interfaces with defense and security agendas. Such a framework must define the limits of dual-use research, provide institutions with clear compliance tools and ensure academic freedom is protected while upholding international ethical standards. Only through transparent and consistent policy can Spain align its research system with both its strategic interests and its democratic values.

\subsection{FONAS statement against undermining the separation of civilian and military research \\
Authors: Thea Riebe, Technical University of Darmstadt, Germany, J\"urgen Altmann, Technical University of Dortmund, Germany}

FONAS (Forschungsverbund Naturwissenschaft, Abr\"ustung und internationale Sicherheit, Research Association for Science, Disarmament and International Security) is the professional organisation of researchers in German-language countries working on questions of disarmament and international security with natural-science, engineering, computer-science or mathematics methods. Reacting to calls for strengthening dual-use and military research at universities, FONAS has issued a statement in February 2025~\cite{FONAS-2025}. In this, FONAS strongly warns against the increasing overlap between civilian and military research. Current political initiatives at both the national and European levels threaten to blur this fundamental boundary, thereby fundamentally altering the nature of science.

FONAS sees this as a significant danger to the independence and transparency of science. {\it The targeted promotion of dual-use technologies creates incentives for the militarization of civilian research institutions and universities}, warns the research network. {\it This not only leads to the deliberate creation of gray areas but also makes it more difficult for researchers to actively choose an exclusively civilian use of their work}. Particularly problematic is the looming weakening or even abolition of Civil Clauses. Without these regulations, the responsibility for ethical research decisions would be unilaterally shifted onto individual scientists, while at the same time, third-party funding and career incentives are increasingly directed toward militarily relevant research fields.

FONAS emphasizes that science thrives on transparency, cooperation, and international exchange. A stronger integration of military interests not only endangers these fundamental principles but also the international attractiveness of Germany as a research location. Stricter security regulations and confidentiality obligations in dual-use projects could exclude foreign researchers and hinder free collaboration in key scientific fields.\\

FONAS therefore calls for:
\begin{itemize}
\item The consistent maintenance and strengthening of the separation between civilian and military research.
\item A clear stance from universities and research institutions against the militarization of science.
\item Strengthening Civil Clauses to ensure transparency and ethical responsibility in research. 
\end{itemize}

\section{Dual-Use Issues}
\subsection{Dual-use and militarization of science and technology \\
Author: J\"urgen Scheffran, Hamburg University}

Civil-military interactions have shaped the history of science, technology and war~\cite{Altmann_2017}. Scientific knowledge and technical know-how became essential preconditions for weapons development and sources of proliferation~\cite{Scheffran2019}. The military was often suggested as a pacemaker of technology development, although the spin-offs were exaggerated. Civilian products are not optimized for military tasks and vice versa. An advantage over opponents is not guaranteed if the technology is available on the international market. The dichotomy between civilian and military technology was pronounced during the East-West conflict when large-scale science and technology became part of the military-industrial complex.

The boundaries eroded after the end of the Cold War. Facing disarmament, lack of public acceptance and budgetary resources, combined with converging demand profiles, supported the dual-use of technologies that have military and civilian applications. Dual-use strategies strengthened civil-military linkages in scientific and technological development and systematically exploited the ambivalence of science and the dual-use of technology~\cite{Scheffran2018,Liebert1994}. A memorandum pointed to the dangers and made proposals for controlling military technology innovation and for redirecting research and technology policy~\cite{ianus-1003}. 

In the new world disorder new crises and conflicts emerged, leading to new civil-military structures and incentives for rearmament, deliberately exploiting civil research for military purposes. The military benefits from research and technologies in the civilian-commercial sector and saving on development costs, shifting research from government to private sectors producing for the mass market. The armament industry pushes for the promotion of dual-use research and the integration of civilian research. 

Taking advantage of its military-technological superiority, the United States and the NATO alliance continued their projections of power and force against competitors in Europe, Russia, China and the Global South and focused on controlling scientific and technological capabilities that may enhance or degrade military capabilities. In the context of the Gulf and Iraq Wars and the terror attacks of September 11, 2001, the quest to control proliferation has increased, as well as the list of potential dual-use technologies, while arms control agreements such as the  Intermediate-Range Nuclear Forces Treaty (INF) Treaty were terminated.

This included controlling exports of {\it sensitive} technologies to {\it critical} countries, in particular, the proliferation of weapons of mass destruction  (nuclear, chemical, or biological) and delivery systems, including related dual-use items which are subject to strict export controls. Since the 1980s the export control focus shifted from an East-West to a North-South context, including technologies on the Trigger List of the London Nuclear Suppliers Club, the Australia Group for chemical weapons, and the Missile Technology Control Regime (MTCR)~\cite{Brauch_1992}.

In recent years, new crises, conflicts and enemy images have promoted growing armaments and expanded security conceptions, creating new justifications for military interventions. The global competition in science and technology has intensified, and is drawn into the arms race, exploiting its dual-use potentials~\cite{Scheffran2019}. The {\it Revolution in Military Affairs} encompasses almost the entire high-tech sector  and draws it into  warfare~\cite{Neuneck2008}. Materials and semiconductors, micro-, nano-, and biotechnologies, nuclear, laser and space, computer and communication technologies, drones and sensors, digitalization and AI are employed not only in civilian products but also in weapons~\cite{Reuter2024}. New wars involve extensive networking, robotization and automation of the battlefields in air, water and ground, in space and cyberspace, with hybrid and information wars on the home front, in social networks and the media~\cite{Springer2018}. Societies are drawn into these wars, connecting civilian and military infrastructures across local and global levels, diffusing clear dividing lines. 

Following the turning point (Zeitenwende) proclaimed  in response to Russia's attack on Ukraine in February 2022, Europe is becoming increasingly militarized~\cite{Scheffran2024}. In order to make society {\it war-ready} by 2029 (as stated by the German defense minister in 2024 \cite{pistorius2024}), Germany is striving to play a {\it leading role}, including key warfare technologies, which is driving the militarization of science and pushing back civilian-oriented science, including repressive measures against critical positions, impairing academic freedom~\cite{Sueddeutsche_2024}. 
A report by the {\it Research and Innovation Expert Commission} ~\cite{efi-2024} recommended that the {\it strict separation} of civilian and military research and development should be fundamentally reconsidered and dissolved where appropriate, in order to consciously promote spillovers and dual-use between military and civilian sectors. Aims for civil-military synergies can also be found in a position paper on research security in changing times of the BMBF~\cite{BMBF-2024}, in the {\it White Paper} \cite{EU-WhitePaper} and the {\it ReArm Europe Plan/Readiness 2030}~\cite{Rearm2030} of the EU commission.

The Coalition Agreement~\cite{Koalitonsvertrag-2025} of the new German government declares that it is {\it committed to removing obstacles that make dual-use research or civil-military research cooperation more difficult}. Promoting future technologies for the armed forces applies to the following areas: {\it satellite systems, artificial intelligence, unmanned (including combat-capable) systems, electronic warfare, cyber, software-defined defense and cloud applications as well as hypersonic systems}, which requires {\it simplified access and in-depth exchange with research institutions, the academic environment, start-ups and industry}.

Against these tendencies, resistance in the research community is emerging, calling for scientific principles such as truth, risk avoidance, peace promotion, social and ecological compatibility. Due to the historical experience with fascism and world wars, there is a widespread rejection of military research and the associated logic of war, secrecy and enemy perception. Freedom of science according to the German Constitution is compatible with human dignity, social responsibility and strengthening democracy. Good research cannot thrive if it is subject to economic, political or military constraints; it needs openness, transparency and critical self-reflection, which requires a civilian orientation.

This is where the Civil Clauses at many universities come in: commitments to teach, learn and conduct research solely for civilian and not military purposes~\cite{Braun2015}. Even if Civil Clauses cannot exclude the military relevance of civilian research in general, they can increase the barriers to military research and strengthen peaceful alternatives. As the Civil Clauses at universities stand in the way of militarization they are increasingly under political pressure. 

In order to escape the logic of war, an ambivalence analysis contributes to transparency at the interface of civil-military research and development and identifies nodes where development paths can be separated on the basis of concrete parameters, as is usual in export control. In doing so, differences between civil and military  applications need to be made clearer rather than blurred, and the social and international framework conditions of decision-making processes should be revealed~\cite{Liebert1994}. 

Challenges of new technologies in complex conflict landscapes require novel approaches of preventive arms control combined with political and legal frameworks that tackle the dual-use problem in the early phases of research and development~\cite{Altmann_1998}.  Preventive arms control, technology assessment and design and Civil Clauses limit dangerous technological developments, influence technological conflicts and create transparency in science. Civil science and education promote international cooperation and solutions to global challenges (against climate crisis, war, displacement and social inequality, and work for just, sustainable and peaceful transformation).

\subsection{Responsibilities of Science Institutions and Scientists}

Science and scientists were always trapped between  applications for public resources to fund scientific research, requests from society providing the financial resources, their own job positions, and ideas for new scientific research.  Science will never be independent of the political and economic situation in which it is embedded.  
Therefore scientists and science institutions cannot deny responsibility for the application of scientific results. A prominent example is the responsibility of scientists during Germany's Nazi period, in their work for developing a nuclear bomb, as well as that of the scientists in the Manhattan project. As a consequence of this responsibility the Einstein-Russell manifesto~\cite{Einstein-Russel-Manifest-1955} was launched, 70 years ago in 1955, where it is written: {\it we urge [the governments], consequently, to find peaceful means for the settlement of all matters of dispute between them.}

In a very similar spirit, the G\"ottingen Eighteen~\cite{manifestoGoettinger18} (among which were M.~Born, O.~Hahn,  M.~v.~Laue, and C.F.~v.~Weizs\"acker) wrote: {\it Our profession is pure science and its application. Conveying our knowledge to young people make us responsible for the consequences of our profession. This is why we cannot remain silent on all political questions.} And they continue: {\it None of the signees would be willing to work on the production, testing of use of nuclear weapons.}

While the development of nuclear weapons was a political decision,  scientists made this development possible. Therefore any participation, directly or indirectly, in military projects falls also under their responsibility. However, not every cooperation with military is against humanistic and peaceful goals, for example participation in arms control as well as in verification of disarmament is at a different level than participation in testing and development material for weapon production. 

Civil Clauses of research institutions help to guide scientists and science institutions towards peaceful research, towards a commitment for a world without war, and especially without nuclear weapons. The discussion on scientific projects must be performed openly, identifying advantages as well as risks, so that everybody is aware of potential consequences and applications. Ethical principles must be part of the education process and be considered as an important part of a universal scientist, in accordance with the words of Major-General E.-P. Nares~\cite{TU-Berlin-1946} at the opening of the TU Berlin, where he argued that every scientist {\it must ask not only "Can I do a good job?" but also: "Will it be put to good use?"} and he continued saying: {\it  And remember that Society is not one nation nor one class of men, but is the whole world and all men and nations in it. }

Civilian scientific projects should have clear international, peacekeeping, bridge-building aspects and all results must be made available to the public in an open-access way. 
In applications for new projects, these aspects should be spelled out clearly as advantages, and risks of  {\it  bad} applications should be considered and evaluated. 
Openness, transparency and internationality are essential, in contrast to secrecy and restriction of access. 
It must be obligatory that projects are followed from an external, historical and societal perspective under the premise of peaceful research, and necessary resources and scientific positions for doing so must be created. \\

We call upon our scientific colleagues and the scientific community
\begin{itemize}
\item to work for using scientific results obtained in public universities and research institutions are used only for peaceful means and for the benefit of humanity, excluding research which supports wars,
\item to take responsibility not only for good scientific research and education but also for the application of scientific results, as part of {\it Good Scientific Practice},
\item to speak out publicly when scientific results are applied and used for non-peaceful purposes,
\item to commit to ethical principles even under the pressure of limited financial funding,
\item to maintain the strict separation of civilian and military research,
\item to uphold the spirit of the CERN Convention to {\it  have no concern with work for military requirements} as a basic principle of Science For Peace.
\end{itemize}

\section{Conclusions}

The discussion about military research in civilian research institutions and public universities has been fueled by the {\it White Paper} and  the  {\it ReArm Europe Plan/Readiness 2030} of the EU commission, as well as position papers by the German governments and the new coalition agreement in Germany. The new proclaimed goal of NATO countries to increase significantly  military spending raises the question of financial consequences for civilian non-military spending. 

In this context the proposal to open civilian research institutions to military research and explicitly support dual-use research is becoming a major threat to universal, peaceful and international research, even at national institutions. In Germany the discussion whether the so-called Civil Clauses at universities and research centers are still adequate was picked up by the DESY directorate in 2024, which immediately resulted in a petition to keep the purely civilian focus of the institute. 

The Science4Peace Forum has initiated a detailed and scientific discussion this topic with a panel discussion of relevant experts in 2024, and this writeup summarizes the many arguments for keeping a strict separation of civilian and military research, while acknowledging that sometimes this separation might be difficult.

We see a dangerous development emerging in science policy, starting from the exclusion of scientists from certain countries from  international projects, claiming that science should play an important role in geopolitical matters and  now explicitly supporting especially dual-use research. Such attempts question the universality and independence of science, criteria that are essential to obtain scientific results for the benefit of humanity and which could help solving the most important problems, which are climate change, ecocide, poverty and wars.

We argue that scientific institutions and individual scientists must take responsibility not only for good scientific results, but also for the application of scientific results for {\it the good} and for the benefit of all people. We call for commitment to ethic principles and to uphold the spirit of the CERN Convention to {\it  have no concern with work for military requirements} as a basic principle of international and universal scientific cooperation.

We argue, that science should never again become involved in the preparation of war, and take seriously the lessons from World War II, lessons which led to the foundation of CERN and the Science4Peace idea.

\vskip 1 cm 
\begin{tolerant}{8000} 
\noindent 
{\bf Acknowledgments.} 
We thank all participants of the {\it Science4Peace Forum} for contributing to the discussions during this Panel. 

\noindent 

\end{tolerant} 
\vskip 0.6cm 

\bibliographystyle{mybibstyle-new}
\raggedright  
\providecommand{\href}[2]{#2}\begingroup\raggedright\endgroup

\end{document}